\documentclass[namedreferences]{solarphysics}

\usepackage[optionalrh,solaromanenum,linksfromyear,natbib]{spr-sola-addons} 
\usepackage{graphicx}                    
\usepackage{amssymb}                    
\usepackage{color}                       
\usepackage{url}                         
\usepackage[pdfborder={0 0 0 },urlcolor=blue,breaklinks]{hyperref}
\usepackage[normalem]{ulem}	
\hypersetup{colorlinks, citecolor=blue, filecolor=black, linkcolor=black, urlcolor=blue}

\makeatletter
\newcommand{\rmnum}[1]{\romannumeral #1}
\newcommand{\Rmnum}[1]{\expandafter\@slowromancap\romannumeral #1@}
\makeatother
\newcommand{\ion}[2]{#1~{\sc \rmnum{#2}}}
\newcommand{\arcsec}{\hbox{$^{\prime\prime}$}}
\newcommand{\degr}{\hbox{$^{\circ}$}}

\newcommand{\corr}[1]{{{#1}}}

\begin{document}

\begin{article}

\begin{opening}

\title{Measuring the Magnetic Field Strength of the Quiet Solar Corona Using ``EIT Waves''}

%
\author{D.M.~\surname{Long}$^{1}$\sep
        D.R.~\surname{Williams}$^{1}$\sep
        S.\surname{R\'{e}gnier}$^{2}$\sep
        L.K.~\surname{Harra}$^{1}$
       }

%
\runningauthor{D.M.~Long et al.}
\runningtitle{Coronal Seismology Using ``EIT Waves''}

%
  \institute{$^{1}$ Mullard Space Science Laboratory, University College London, Holmbury St. Mary, Dorking, Surrey, RH5 6NT, UK.
                     Email: \url{david.long@ucl.ac.uk} \\
                 $^{2}$ Jeremiah Horrocks Institute, University of Central Lancashire, Preston, Lancashire, PR1 2HE, UK     
             }

\begin{abstract}
Variations in the propagation of globally-propagating disturbances (commonly called ``EIT waves'') through the low solar corona offer a unique opportunity to probe the plasma parameters of the solar atmosphere. Here, high-cadence observations of two ``EIT wave'' events taken using the Atmospheric Imaging Assembly (AIA) instrument onboard the \emph{Solar Dynamics Observatory} (SDO) are combined with spectroscopic measurements from the Extreme ultraviolet Imaging Spectrometer (EIS) onboard the \emph{Hinode} spacecraft and used to examine the variability of the quiet coronal magnetic-field strength. The combination of pulse kinematics from \corr{SDO}/AIA and plasma density from \emph{Hinode}/EIS is used to show that the magnetic-field strength is in the range \corr{$\approx$}~2\,--\,6~G in the quiet corona. The magnetic-field estimates are then used to determine the height of the pulse, allowing a direct comparison with theoretical values obtained from magnetic-field measurements from the Helioseismic and Magnetic Imager (HMI) onboard \corr{SDO} using PFSS and local-domain extrapolations. While local-scale extrapolations predict heights inconsistent with prior measurements, the agreement between observations and the PFSS model indicates that ``EIT waves'' are a global phenomenon influenced by global-scale magnetic field.
\end{abstract}

%

\end{opening}

%
\section{Introduction}
\label{S:intro} 

Although the solar corona is dominated by the Sun's magnetic field, accurately determining its strength continues to be a difficult task. Estimates may be obtained in a small number of long-wavelength (forbidden) emission lines by measuring their Zeeman splitting \citep[\emph{cf}.][]{Lin:2004ab} or by using the Hanle effect \citep[\emph{cf}.][]{Raouafi:2002ab}, but these are generally obtained near active regions where the magnetic-field strength is high and very strong lines in the near infrared may be used \citep[such as, \emph{e.g.,} \corr{\ion{Fe}{13}}~$\lambda$10\,747;][]{Lin:2000ab}. The strength of the coronal magnetic field can also be derived from the gyro-resonance emission in radio wavelengths \citep[\emph{e.g.,}][]{White:1997ab}. However, the optically thin emission lines and weak magnetic-field strength above quiet-Sun regions mean that these techniques are typically not suited to measuring magnetic field strength there. An alternative approach is to infer the coronal magnetic-field strength and other plasma parameters by examining how the properties of waves change with propagation: a technique called \emph{coronal seismology} \citep{Uchida:1970ab,Roberts:1984ab}.

From their initial discovery in images from the \corr{\emph{Extreme ultraviolet Imaging Telescope}} \citep[EIT:][]{Delaboudiniere:1995ab} \corr{onboard the \emph{SOlar and Heliospheric Observatory} (SOHO) spacecraft}, large-scale coronal disturbances \citep[commonly called ``EIT waves'':][]{Dere:1997ab,Moses:1997ab,Thompson:1998ab} have been suggested as possible probes for studying the plasma parameters of the low corona \citep[\emph{e.g.,}][]{Ballai:2003ab,Ballai:2004ab}. These pulses are quite fast, with typical velocities of \corr{$\approx$} 200\,--\,400~km~s$^{-1}$ \corr{measured using data from \corr{SOHO}/EIT} \citep{Thompson:2009ab} and have been observed to exhibit deceleration \citep{Long:2008ab,Warmuth:2004a} and pulse broadening \citep{Long:2011a,Muhr:2011ab} with propagation: features consistent with the propagation of a fast-mode magnetohydrodynamic (MHD) wave through a randomly structured medium \citep[\emph{e.g.,}][]{Murawski:2001ab}. The pulses appear to expand isotropically through \corr{quiet-Sun regions}, \corr{although they do tend to avoid regions of lower and higher density such as} coronal holes and active regions \citep{Thompson:1999ab}, a property which makes them ideal for determining the nature of the quiet corona.

However, the use of ``EIT waves'' to probe the quiet corona is predicated on the interpretation of these disturbances as MHD waves, which is not a simple issue. There are currently two competing physical interpretations for these disturbances. The first uses MHD wave theory to explain the phenomenon \citep[\emph{e.g.,}][]{Long:2008ab,Veronig:2010ab,Wang:2000ab,Kienreich:2012ab,Patsourakos:2009ab} with observations of wave properties such as reflection and refraction at coronal-hole boundaries \citep{Gopal:2009ab} supporting this approach. An alternative interpretation visualises the pulse as a ``pseudo-wave'' resulting from the restructuring of the global magnetic field during the eruption of a coronal mass ejection \citep[CME: \emph{e.g.,}][]{Delannee:2008ab,Schrijver:2011ab,Attrill:2007ab}. In this scenario, the bright feature observed as the ``EIT wave'' is due to Joule heating or small-scale magnetic reconnection as the erupting CME passes out of the low corona. 

An alternative explanation for ``EIT waves'' combines both the wave and ``pseudo-wave'' theories to interpret this phenomenon as consisting of both a fast-mode wave initially driven by the erupting CME and a slower brightening due to reconfiguration of the magnetic field. This form was originally posited by \citet{Chen:2002ab} and has been expanded in simulations performed by \citet{Chen:2005ab,Cohen:2009ab} and \citet{Downs:2011ab,Downs:2012ab}. There has also been some evidence of two distinct fronts in observations \citep[\emph{e.g.,}][]{Zhukov:2004ab,Chen:2011ab}. Recent statistical analysis of a large sample of ``EIT waves'' performed by \citet{Warmuth:2011ab} suggests that there may be three distinct classes of ``EIT wave''. Class~1 pulses are initially fast waves that exhibit pronounced deceleration, Class~2 pulses are waves with moderate and almost constant velocities, while Class~3 pulses exhibit erratic kinematic behaviour and are most likely explained as pseudo-waves.

``EIT waves'' are traditionally identified and analysed using data from imagers such as \corr{SOHO}/EIT \citep{Thompson:1999ab}, the \emph{Transition Region And Coronal Explorer}  \citep[TRACE, \emph{e.g.,}][]{Wills-Davey:1999ab}, the Extreme UltraViolet Imager (EUVI) onboard the \emph{Solar Terrestrial Relations Observatory} \citep[STEREO, \emph{e.g.,}][]{Long:2008ab}, and the Atmospheric Imaging Array (AIA) onboard the \emph{Solar Dynamics Observatory} \citep[SDO, \emph{e.g.,}][]{Liu:2010ab}, as these instruments allow easy identification of the disturbance within a large field-of-view. However, a more detailed understanding of these disturbances requires the use of spectroscopic instruments as these allow their true physical nature to be investigated. This approach is hindered by the fact that these instruments generally have a restricted field-of-view, making observations of ``EIT Waves'' rare. Despite this, several events have been observed using the \corr{\emph{Extreme ultraviolet Imaging Spectrometer}} \citep[EIS:][]{Culhane:2007ab} onboard the \emph{Hinode} spacecraft \citep{Kosugi:2007ab}. 

As rare cases of ``EIT waves'' being observed by spectroscopic instruments, both of the events discussed here have previously been studied by other authors. The event from 12~June~2010 was analysed by \citet{Chen:2011bc}, who found a significant change in the magnetic topology as the nearly circular pulse passed over an upflow region near a magnetic bipole. This led the authors to suggest that the event may be best explained using the magnetic field-stretching model proposed by \citet{Chen:2005ab}. The 16~February~2011 event was one of a series of eruptions from AR 11158 over the course of several days and the active region was the subject of a specialist \emph{Hinode} Observing Plan (HOP) for studying ``EIT waves''. The eruption has been studied by both \citet{Harra:2011ab} and \citet{Veronig:2011ab} who found clear downward bulk motion towards the chromosphere at the pulse, followed by a later upward motion behind the pulse. The kinematics measured by \corr{\emph{Hinode}}/EIS also matched those using imagers, suggesting that ``EIT waves'' may be best interpreted as MHD waves propagating through the low corona.

In this article, we examine \corr{those} two ``EIT wave'' events observed by both \emph{Hinode}/EIS and \corr{SDO}/AIA. The spectroscopic observations from \corr{\emph{Hinode}}/EIS are used to determine the density of the plasma through which the pulse is propagating, and this information is combined with the kinematics obtained from \corr{SDO}/AIA, allowing the magnetic-field strength of the quiet corona to be estimated and compared with theoretical predictions from multiple extrapolated-field models. The observations of both events studied are outlined in Section~\ref{S:data}, with the analysis of these observations from both \corr{SDO}/AIA and \corr{\emph{Hinode}}/EIS discussed in Section~\ref{S:methods}. The results of this analysis are presented in Section~\ref{S:results} before being compared to extrapolated-field estimates in Section~\ref{S:mag_field}. Finally, some conclusions are outlined in Section~\ref{S:concs}.

\section{Observations and Data Analysis}
\label{S:data} 

The events studied here were quite similar. The 12~June~2010 eruption from active region (AR) NOAA~11081 was associated with a GOES M2.0 flare, which began at 00:30~UT and peaked at 00:57~UT, while the eruption on 16~February~2011 from AR NOAA~11158 had an associated M1.6 flare (starting at 14:19~UT, peaking at 14:25~UT). Both events were also associated with Type~\Rmnum{2} radio bursts, while CMEs were also identified using the Coordinated Data Analysis Workshops (CDAW) catalogue (\href{http://cdaw.gsfc.nasa.gov/CME_list/}{http://cdaw.gsfc.nasa.gov/CME\_list/}) for the 12~June~2010 event and by the instrument team for the Cor-1 coronagraph (\href{http://cor1.gsfc.nasa.gov/catalog/}{http://cor1.gsfc.nasa.gov/catalog/}), part of the Sun Earth Connection Coronal and Heliospheric Investigation (SECCHI) instrument package onboard the STEREO spacecraft for the 16~February~2011 event.

\begin{figure*}[!t]
\centering
\includegraphics[width=0.9\textwidth,clip=,trim=0 30 0 0]{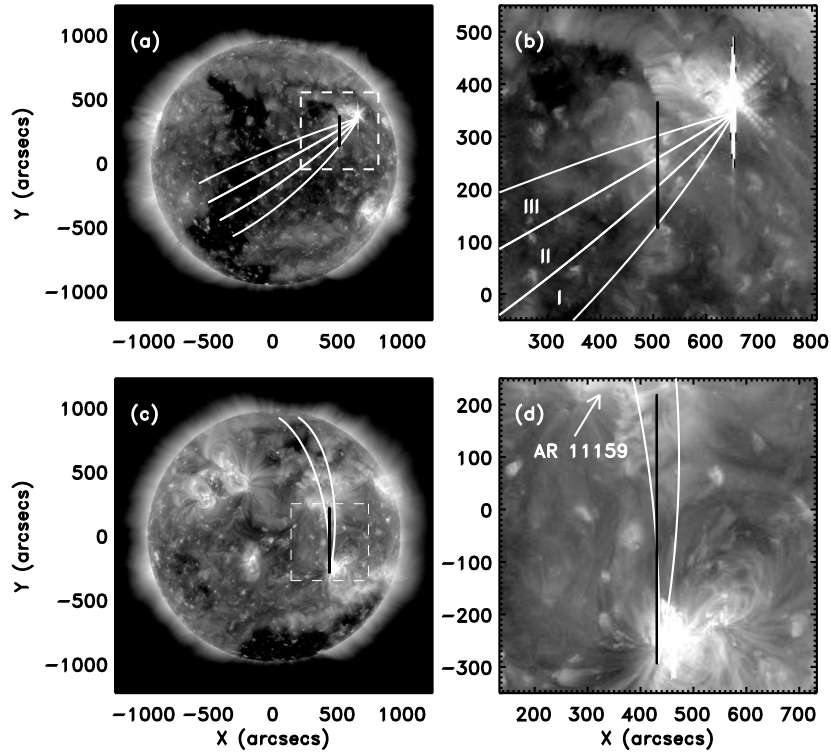}
\caption{SDO/AIA 193~\AA\ images showing the events from 12~June~2010 (\corr{\emph{t}} = 00:58:30~UT; full-disk in panel (a) and zoomed-in in panel (b)) and 16~February~2011 (\corr{\emph{t}} = 14:25:31~UT; full-disk in panel (c) and zoomed-in in panel (d)). The great-circle arc sectors used to determine the kinematics of the pulses are shown bounded in white, while the \corr{\emph{Hinode}}/EIS slit positions are in black in all panels. \corr{AR~11159, which lay to the North of the erupting AR~11158 for the 16~February~2011 event, is indicated in panel (d).} The arcs in panel (b) are labelled \Rmnum{1}, \Rmnum{2}, and \Rmnum{3} for easier identification throughout the text.\corr{The variation of the zoomed-in regions with time is shown in the running-difference movies~1 and 2 in the electronic supplementary material.}}
\label{fig:overview}
\end{figure*}

The \corr{\emph{Hinode}}/EIS data for these events were taken from two separate observation programmes, and consequently measured different sets of emission lines. The data for the 12~June~2010 event were taken using EIS study~387, which was designed to study the asymmetry of transition region emission lines. This produced a set of 12 observations of \corr{$\approx$}~five~minute duration taken in a region of the quiet Sun adjacent to the erupting active region (as shown in panels (a) and (b) of Figure~\ref{fig:overview}). The data for the 16~February~2011 event were taken using HOP~180, which was a co-ordinated programme designed to study the plasma properties of ``EIT waves''. This data-set consists of a single sit-and-stare raster observing both the edge of the erupting active region and the adjacent quiet Sun (see panels (c) and (d) of Figure~\ref{fig:overview}) and lasting \corr{$\approx$}~30~minutes.

Full-disk images (0.6\corr{\arcsec}~pixel$^{-1}$) from the AIA instrument \citep{Lemen:2012ab} onboard the SDO spacecraft \citep{Pesnell:2012ab} were used to determine the kinematics and morphological evolution of both ``EIT waves''. The \corr{SDO}/AIA data used here were taken from the 193~\AA\ passband (which is sensitive to coronal plasma at \corr{$\approx$}~1-2~MK) as it provided the clearest observations of the pulse. The kinematic properties of both pulses were determined using the semi-automated intensity-profile technique utilised by \citet{Long:2011a} and \citet{Long:2011b}. This is outlined in more detail in Section~\ref{SS:methods_aia}.

\section{Methods}
\label{S:methods} 

Since our data are from two distinct instruments, each measuring different properties of the observed phenomenon, analysis of the observations from both \corr{\emph{Hinode}}/EIS and \corr{SDO}/AIA was a two-step process. The full-disk images from \corr{SDO}/AIA were primarily used to examine the kinematics of the disturbances, as well as to allow some comparison with the magnetic field extrapolations obtained from the \corr{\emph{Helioseismic and Magnetic Imager}} \citep[HMI:][]{Scherrer:2012ab} onboard \corr{SDO}. The spectra obtained from \corr{\emph{Hinode}}/EIS were used to determine the \corr{number} density of the quiet-Sun region being examined using several density-sensitive line ratios.

\subsection{\corr{SDO}/AIA analysis}
\label{SS:methods_aia}

The kinematics of the two pulses studied were determined using the semi-automated intensity-profile technique promulgated by \citet{Long:2011a} and \citet{Long:2011b}. This technique uses percentage base-difference intensity \citep[PBD:][]{Wills-Davey:1999ab} images to identify the disturbance, with each image derotated to the same pre-event time for each event. A series of 36 arcs of \corr{$\approx$}~10\corr{\degr} width radiating from a source point are used to create intensity profiles by averaging the PBD intensity across each arc sector in annuli of increasing radii and 1\corr{\degr} width on the spherical surface. \corr{As this algorithm is designed to operate automatically with minimal user input, the source point from which to measure the distance of the pulse is taken as the position of the flare defined by the Heliophysics Event Knowledgebase (\href{https://www.lmsal/hek/isolsearch/isolsearch.html}{https://www.lmsal/hek/isolsearch/isolsearch.html}; note that this is used to ensure consistency between events, but does not imply that the flare is the physical source of the wave).} This creates a set of 36 intensity profiles for each image, with the \corr{mean and standard deviation of the PBD intensity values across the annulus taken as the intensity and associated error for that point on the profile}. 

\corr{Once the intensity profiles had been obtained, the intensity peak corresponding to the flare, the propagating pulse, and any associated ``stationary brightenings'' were identified automatically by the algorithm for each arc, with a Gaussian model used to fit the position of the peak intensity of each feature individually. The pulse was then identified through its motion, with the pulse position, height, and width given by the centroid, peak, and full width at half maximum (FWHM) respectively \citep[as ``EIT waves'' display a Gaussian cross-section, see][]{Wills-Davey:1999ab}. The errors associated with each parameter were obtained from the error on the fit to the intensity profile. This approach was used as it operates automatically, thus minimising user bias; there is no user input into the algorithm since the source of the arcs and start time of the analysis were both obtained from the identification of the flare defined by the HEK. In addition, the algorithm is designed to compare features between arc sectors, ensuring their accurate identification.} The pulse position and width were therefore obtained with respect to time for each arc, allowing the kinematics and temporal behaviour of the pulse to be determined. The kinematics of the pulse were derived by fitting the pulse position with time using a quadratic equation of the form
\begin{equation}
r(t) = r_0 + v_0 t + \frac{1}{2}a t^2,
\end{equation}
where $r_0$ is the initial pulse position, $v_0$ is the initial pulse velocity, and $a$ is the acceleration of the pulse. Broadening of the pulse was identified by fitting the variation in pulse width with time using a linear function. \corr{Note that to compare the \corr{SDO}/AIA kinematics for these events directly with the \corr{\emph{Hinode}}/EIS observations, only the three arcs that intersected the \corr{\emph{Hinode}}/EIS slit were used for the 12~June~2010 event, while just one arc was suitable for the 16~February~2011 event. This is shown in Figure~\ref{fig:overview}.}

\begin{figure*}[!t]
\centering
\includegraphics[width=0.95\textwidth,clip=,trim=20 20 0 0,angle = 0]{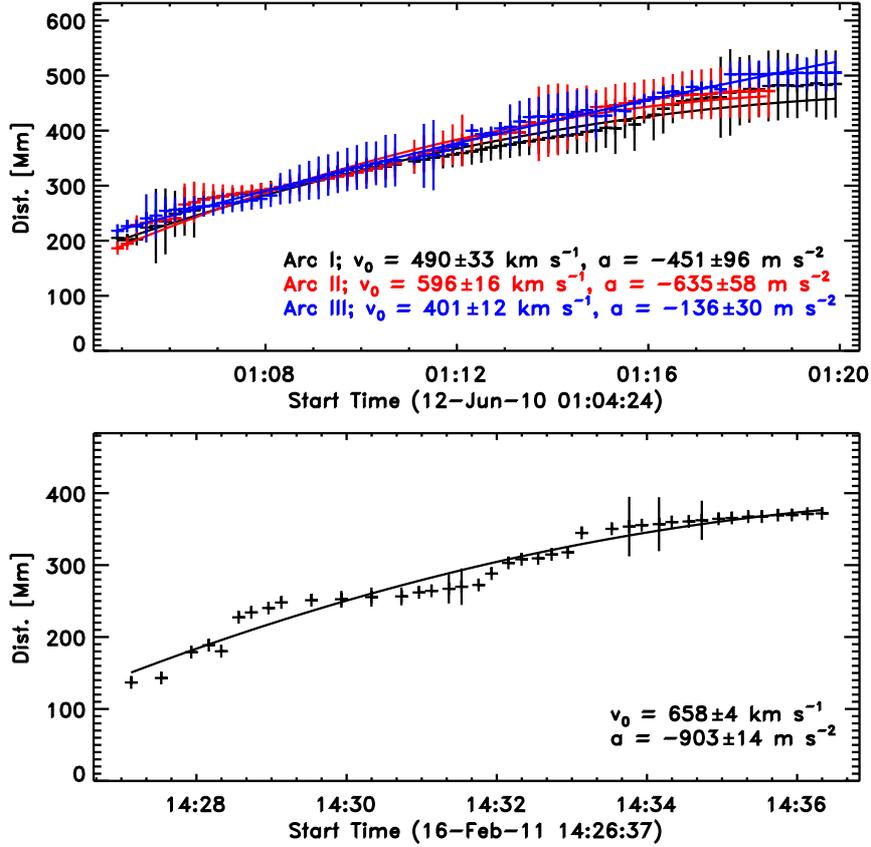}
\caption{Time-distance plots for the events of 12~June~2010 (top) and 16~February~2011 (bottom). The kinematics derived for each of the arcs \Rmnum{1}, \Rmnum{2}, and \Rmnum{3} for the 12~June~2010 and 16~February~2011 ``EIT waves'' are given in the respective legends.}\label{fig:kins}
\end{figure*}

The resulting distance-time plots produced by this analysis are shown in Figure~\ref{fig:kins}. The kinematics derived for each arc were obtained using a residual resampling bootstrapping approach \citep[see, \emph{e.g.},][for more details]{Long:2011a}. This technique was chosen to determine the pulse kinematics as it provides a statistically significant estimate of the errors associated with the derived kinematic values by allowing each parameter to be characterised by a distribution.

The 12~June~2010 pulse was observed by \corr{\emph{Hinode}}/EIS with its slit positioned approximately perpendicularly to the direction of the pulse's propagation (see Figure~\ref{fig:overview}). As a result, it was possible to measure the pulse kinematics where three different arc sectors cross the slit; these measurements are shown in the upper panel of Figure~\ref{fig:kins}. It is clear that while the distance travelled by the pulse in each arc sector is comparable, the resulting pulse kinematics are different for each arc sector. This indicates that the pulse does not propagate isotropically, and \corr{it is therefore of interest to examine whether the differences in kinematics are driven by differences in the local or global plasma conditions, where \corr{\emph{Hinode}}/EIS may be used to study the local variations and \corr{SDO}/AIA may be used to study the variations in global propagation.} 

The geometry of the 16~February~2011 event was slightly different, in that the pulse propagation direction was almost parallel to the \corr{\emph{Hinode}}/EIS slit (see Figure~\ref{fig:overview}). Only one arc sector was therefore required to determine the kinematics of the pulse. The resulting kinematics, shown in the bottom panel of Figure~\ref{fig:kins}, indicate a slightly higher initial velocity and much stronger deceleration. This suggests that while the initial driver of the pulse may have been comparable to that of the 12~June~2010 event, the free propagation of the 16~February~2011 pulse was subject to more resistance.

\subsection{\corr{\emph{Hinode}}/EIS analysis}
\label{SS:methods_eis}

Analysis of the \corr{\emph{Hinode}}/EIS observations for the events studied was complicated by the fact that the data were taken from two distinct observing plans that were designed with different scientific goals in mind. Despite this, both sets of observations included density-sensitive line ratios that can be used to estimate the \corr{electron number-}density variation of the low solar corona through which each pulse propagated. 

Two density-sensitive line ratios were included in the 12~June~2010 observations: the \corr{\ion{Si}{10}}~$\lambda$258.37/261.04 and \corr{\ion{Fe}{14}}~$\lambda$264.78/274.20 ratios, which are sensitive to \corr{electron number}-densities of \corr{log$_{10}(n_{\textrm{e}}$)~$\approx$}~(8\,--\,10) and \corr{log$_{10}(n_{\textrm{e}}$)~$\approx$}~(9\,--\,11)~cm$^{-3}$ respectively \citep{Young:2007ab}. This combination meant that the coronal density could be determined over a temperature range of \corr{log$_{10}(T)$ $\approx$} (6.1\,--\,6.3)~K \citep{Mazzotta:1998ab}, which covers the peak emission temperature of the 193~\AA\ passband observed by \corr{SDO}/AIA (\corr{log$_{10}(T)$~$\approx$}~6.1~K). The 16~February~2011 event was observed using a more targeted EIS observing plan, and therefore the data include four density-sensitive line ratios. These were the \corr{\ion{Fe}{12}}~$\lambda$186.88/195.12, \corr{\ion{Fe}{13}}~$\lambda$196.54/202.04, \corr{\ion{Fe}{13}~$\lambda$203.82/202.04} and \corr{\ion{Mg}{7}}~$\lambda$278.39/280.75 line ratios, corresponding to a \corr{range of densities} of \corr{log$_{10}(n_{\textrm{e}}$)~$\approx$}~(8\,--\,11)~cm$^{-3}$ and a temperature range of \corr{log$_{10}(T)~\approx$}~(5.8\,--\,6.2)~K.

\begin{figure*}[!t]
\centering
\includegraphics[width=0.85\textwidth,clip=,trim=0 0 0 0,angle = 0]{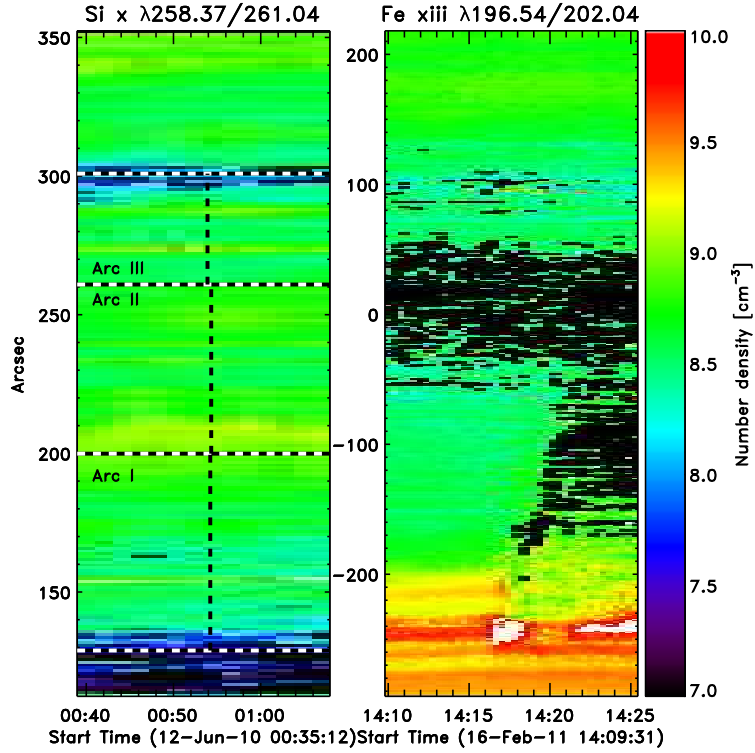}
\caption{\corr{Variation in number density with time calculated for the 12~June~2010 event (\ion{Si}{10}~$\lambda$258.37/261.04 ratio; left panel) and 16~February~2011 event (\ion{Fe}{13}~$\lambda$196.54/202.04 ratio; right panel). The sections delineated by the horizontal lines in the left panel correspond to arc sectors (from bottom to top) \Rmnum{1}, \Rmnum{2}, and \Rmnum{3} respectively, with the vertical dashed lines indicating the time in each arc sector at which the pulse passed through the slit.}}
\label{fig:dens_imgs}
\end{figure*}

For both events, the density-sensitive line ratios were obtained first by averaging the measured intensity in \corr{time} to obtain a one-dimensional intensity profile \corr{along the \corr{\emph{Hinode}}/EIS slit}. The \corr{number}-densities were then calculated using the $\sf{eis\_density.pro}$ routine contained within the SolarSoftWare (SSW) software package, producing a one-dimensional density profile along the slit for both events studied. Although multiple line ratios were available for each event, only the \corr{\ion{Si}{10}~$\lambda$258.37/261.04} and \corr{\ion{Fe}{13}~$\lambda$196.54/202.04} line ratios were used in this analysis \corr{for the 12~June~2010 and 16~February~2011 events respectively} as they were most sensitive to \corr{variations in the range log$_{10}(n_{\textrm{e}}$)~$\approx$~(8\,--\,10)~cm$^{-3}$.  The variation in density with time is shown in Figure~\ref{fig:dens_imgs} for both events studied.} The resulting density profiles are shown \corr{in the upper panels of} Figures~\ref{fig:eis_20100612} and \ref{fig:eis_20110216} for each observed event. 

\section{Results}
\label{S:results} 

The derived kinematic parameters of the pulses studied here are shown in Figure~\ref{fig:kins} to be quite high (with initial velocities of $\approx$~\corr{496}~km~s$^{-1}$ and \corr{658}~km~s$^{-1}$ respectively). These values are higher than the typically observed pulse velocities reported by \citet{Thompson:2009ab}, but are consistent with other measurements using data from \corr{SDO}/AIA reported by \corr{\emph{e.g.}, }\citet{Zheng:2012ab} and \citet{Olmedo:2012ab}. In addition, both events exhibited significant deceleration with values of $a = -$\corr{(136 to 635)}~m~s$^{-2}$ and $-$\corr{903}~m~s$^{-2}$ for the 12~June~2010 and 16~February~2012 events respectively. The higher pulse velocity and significant deceleration observed here are consistent with the Class~1 ``EIT wave'' classification proposed by \citet{Warmuth:2011ab}. According to this classification system, ``EIT waves'' exhibiting a high initial velocity \citep[i.e., $>$325~km~s$^{-1}$:][]{Warmuth:2011ab} and a resulting strong deceleration (such that the final pulse velocity is \corr{$\approx$}~200\,--\,300~km~s$^{-1}$) is thought to correspond to fast MHD wave modes, with the result that the events reported here are interpreted as such.

By interpreting these phenomena as fast-mode MHD waves \corr{\citep[\emph{e.g.},][]{Priest:1987}}, it is possible to examine their kinematics using the equation
\begin{equation}\label{eqn:v_fast_mode}
v_{\textrm{fm}} = \sqrt{v_\textrm{A}^2 + c_\textrm{s}^2},
\end{equation}
where the Alfv\'{e}n speed and sound speed are defined as $v_\textrm{A} = B/(4 \pi n m)^{1/2}$ and $c_\textrm{s} = (\gamma k T/m)^{1/2}$ respectively and the propagation of the pulse is approximately perpendicular to the coronal magnetic field. The magnetic-field strength is defined by $B$, $n$ is the particle \corr{number} density, $m$ is the proton mass, $\gamma$ is the adiabatic index (typically $5/3$), $k_\textrm{B}$ is the Boltzmann constant and $T$ refers to the \corr{peak emission temperature of the density sensitive lines used \citep[see Section~\ref{SS:methods_eis} and][]{Mazzotta:1998ab}}. This equation can then be rewritten in terms of the magnetic-field strength $B$ as,
\begin{equation}\label{eqn:b_field}
B = \sqrt{4 \pi n (m v^2_{\textrm{fm}} - \gamma k_\textrm{B} T)}.
\end{equation}
This approach is discussed in more detail by \citet{Long:2011b} and \citet{West:2011ab} and may be used to estimate the magnetic field strength, given the plasma density and pulse velocity.

\begin{figure*}[!t]
\centering
\includegraphics[width=0.9\textwidth,clip=,trim=0 30 0 0,angle = 0]{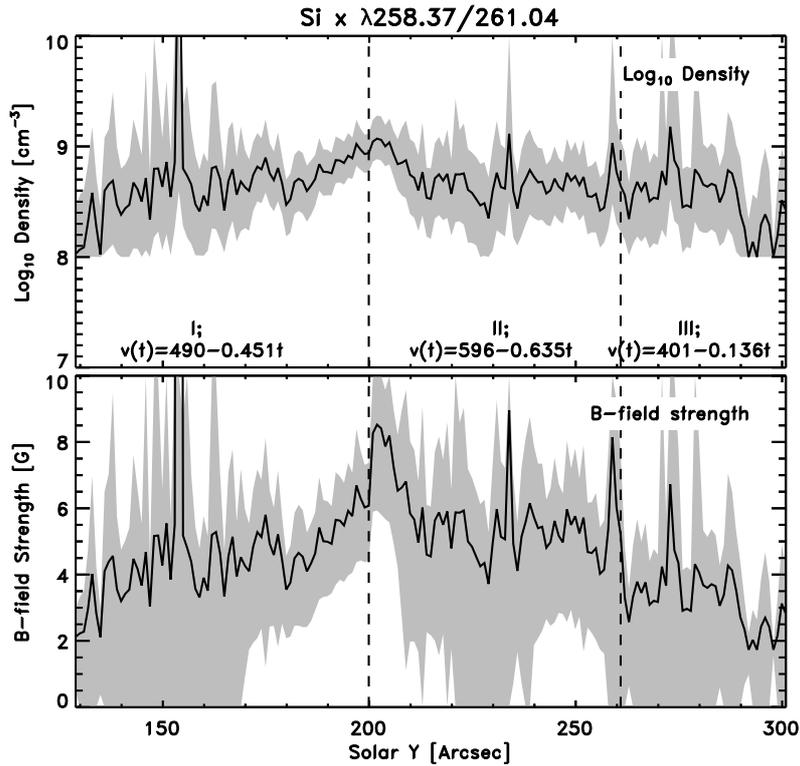}
\caption{Variation in \corr{number} density \corr{(top panel)} and magnetic-field strength \corr{(bottom panel)} with position for the 12~June~2010 event. The sections delineated by the vertical lines correspond to arc sectors (from left to right) \Rmnum{1}, \Rmnum{2}, and \Rmnum{3} respectively \corr{while the error associated with each parameter is indicated by the grey shaded region}. The density variation was determined using the \corr{\ion{Si}{10}~$\lambda$258.37/261.04} density-sensitive line ratio.}\label{fig:eis_20100612}
\end{figure*}

\begin{figure*}[!t]
\centering
\includegraphics[width=\textwidth,clip=,trim=20 0 0 0,angle = 0]{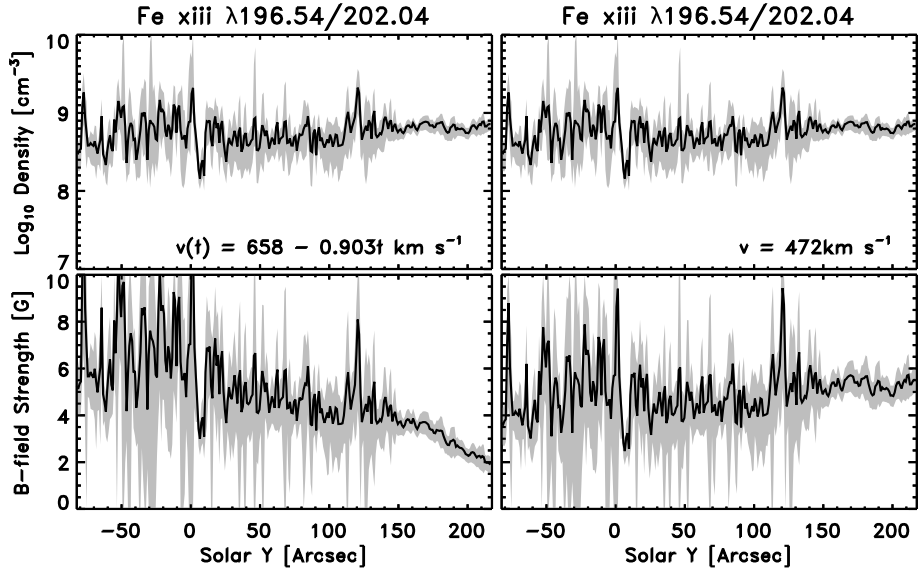}
\caption{Variation in \corr{number} density \corr{(top row)}. Note that both panels are the same in order to calculate magnetic-field strength \corr{(bottom row)} with position for the 16~February~2011 event using a \corr{decelerating pulse velocity (left) and constant pulse velocity (right)}. \corr{The error associated with both the number density and magnetic field strength is indicated by the grey shaded region.} The density variation was determined using the \corr{\ion{Fe}{13}~$\lambda$196.54/202.04} density-sensitive line ratio.}\label{fig:eis_20110216}
\end{figure*}

Equation~(\ref{eqn:b_field}) was used to estimate the quiet coronal magnetic-field strength for both of the events presented here. For the 12~June~2010 event, the distance of the \corr{\emph{Hinode}}/EIS slit from the pulse source was measured for each of the arc sectors studied, and the pulse kinematics were determined for this distance. Resulting velocities of \corr{$\approx$~469}~km~s$^{-1}$, \corr{$\approx$~558}~km~s$^{-1}$, and \corr{$\approx$~400}~km~s$^{-1}$ were estimated for arcs~\Rmnum{1}, \Rmnum{2}, and \Rmnum{3} respectively. These values were combined with the \corr{number} density profile shown in \corr{the upper panel of} Figure~\ref{fig:eis_20100612} to produce the magnetic field strength in the range 2\,--\,6~G, shown in \corr{the lower panel} in the same figure. The sections delineated by the vertical black lines correspond to (from left to right) arc sectors \Rmnum{1}, \Rmnum{2} and \Rmnum{3} respectively.

The slightly different geometry of the 16~February~2011 event meant that the pulse propagated along the \corr{\emph{Hinode}}/EIS slit with a velocity shown in the bottom panel of Figure~\ref{fig:kins}. Two kinematic models for the pulse were examined for this event: a pulse that propagated at a constant velocity of \corr{$\approx$~472}~km~s$^{-1}$, obtained by taking the mean velocity of the pulse across the \corr{\emph{Hinode}}/EIS field-of-view; and a decelerating pulse with a velocity given by\corr{
\begin{equation}
v = 658 - 0.903t\textrm{~km~s$^{-1}$}.
\end{equation} }
These velocity values were combined with the \corr{number} density profile obtained for this event \corr{(top panels of Figure~\ref{fig:eis_20110216})} to produce the magnetic field strength profiles shown in \corr{the bottom panels of} Figure~\ref{fig:eis_20110216}, with the \corr{right panels} showing the results for constant velocity and the \corr{left panels} showing the results for variable velocity. It is clear that while both profiles are similar, the variation in pulse velocity does affect the derived magnetic-field strength. When a constant velocity is assumed, $B$ is in the range 2\,--\,6~G; where deceleration is considered, the range of $B$ increases to 1.5\,--\,10~G, and the difference in results is most easily seen in the upper part of the \corr{\emph{Hinode}}/EIS slit (100\,--\,220\corr{\arcsec}).

These observations suggest \corr{that the magnetic field strength of the quiet corona exhibits some variability with position, ranging here between $\approx$~0\,--\,10~G (within errors)}. Previous estimates of the quiet coronal magnetic-field strength \corr{using ``EIT Waves''} were constrained by the use of lower cadence images from \corr{STEREO}/EUVI or by the use of general estimates of the coronal \corr{number} density rather than precise measurements using \emph{Hinode}/EIS. This was reflected in the resulting general estimates, with values of 0.7$\pm$0.7~G and \corr{$\approx$}~1\,--\,2~G returned by \citet{West:2011ab} and \citet{Long:2011b} respectively. The higher values of magnetic field strength here may reflect the use of measured rather than canonical values of density, in combination with higher-cadence data that better capture the kinematic properties of the pulse.

\section{Magnetic Field Extrapolation}
\label{S:mag_field}

These measurements of the coronal magnetic-field strength invite further investigation and comparison with theoretical models. A detailed examination of the magnetic-field configuration in the vicinity of the \emph{Hinode}/EIS slit was therefore undertaken to try and understand them. Two different methods were used to examine the magnetic field; a potential field source surface (PFSS) model giving a potential-field configuration for the full Sun (discussed in Section~\ref{SS:pfss}), and a potential-field extrapolation in a Cartesian box above the quiet Sun (discussed in Section~\ref{SS:qs_model}). Given the magnetic-field strength estimated using the propagation of the ``EIT wave'' and shown in Figures~\ref{fig:eis_20100612} and \ref{fig:eis_20110216}, the methods below are used to derive the height at which this field strength is most probable.

\subsection{PFSS model}
\label{SS:pfss}

\begin{figure*}
\centering
\includegraphics[width=1\linewidth,trim=0 25 0 0, clip=]{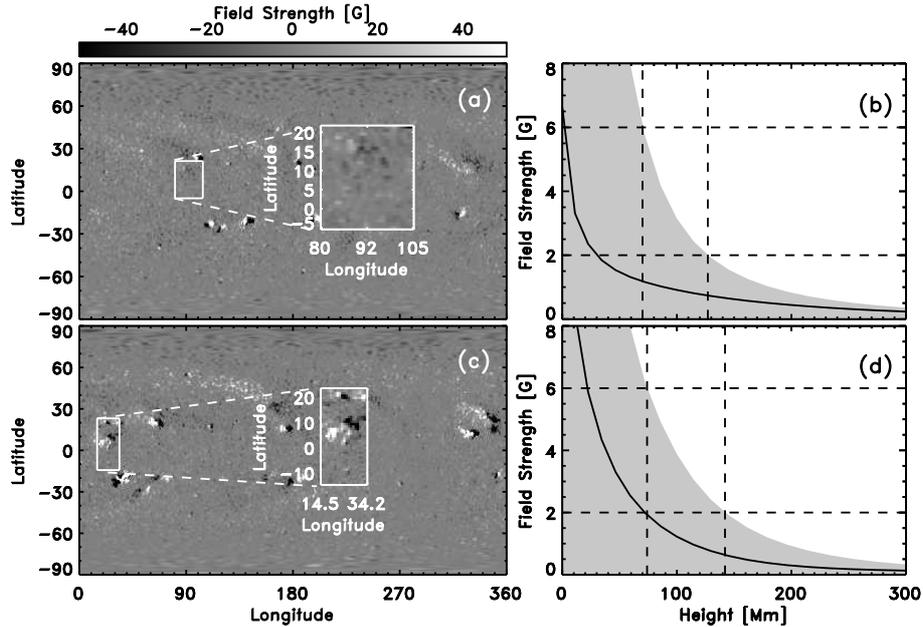}
\caption{\corr{Synoptic maps from 12~June~2010 (panel (a)) and 16~February~2011 (panel (c))} used as a boundary condition for the PFSS model. The white box \corr{and enlarged section indicate} the area in which the average magnetic field strength is computed. \corr{The corresponding average field strength as a function of height in the corona is shown in panel (b) for the event of 12~June~2010 and panel (d) for the event of 16~February~2011}. The grey area shows the spread of the magnetic-field strength. The vertical dashed lines indicate the height at which a maximum field strength of 6~G and 2~G is achieved (horizontal lines).}
\label{fig:b_pfss}
\end{figure*}

The PFSS model developed by \citet{alt69}, \cite{sch69}, and \citet{sch03} was used to investigate the magnetic field strength associated with the observed ``EIT waves''. This model has the advantage of requiring only the distribution of the radial magnetic field on the photosphere as a boundary condition. Synoptic maps from \corr{SDO}/HMI were used to determine the photospheric magnetic field for each event, using the maps at 00:04 UT for the 12~June~2010 event (\corr{Figure~\ref{fig:b_pfss}(a)}), and at 12:04 UT for the 16~February~2011 event (\corr{Figure~\ref{fig:b_pfss}(c)}). In both cases, the spatial resolution of the maps is \corr{1\degr}, while the source surface is located at 2.5~R$_{\odot}$ and assumes that the magnetic field is radial above this height. \corr{The PFSS model provides the magnetic field for each pixel of the domain.}

A spherical-co-ordinate wedge \corr{($\approx23^{\circ}\times26^{\circ}$ for the 12~June~2010 event and $\approx19^{\circ}\times37^{\circ}$ for the 16~February~2011 event)} that included the \emph{Hinode}/EIS slit \corr{was then extracted. This} is shown by the white rectangle \corr{and the zoomed in section of panels (a) and (c) of Figure~\ref{fig:b_pfss}}. 

Figure~\ref{fig:b_pfss}(b) shows the variation of the average $B$ field with height for the 12~June~2010 event. The grey area defines the distribution of $B$ delineated by the maximum of $B$ at a given height. The vertical dashed lines indicate the \corr{heights at which $B$ drops below 6~G and 2~G respectively, implying that the height range of this event lies} between 70 and 128~Mm. Similarly for the 16~February~2011 event, Figure~\ref{fig:b_pfss}(d) indicates that the heights at which $B$ is lower than 6~G and 2~G are 72~Mm and 131~Mm respectively.

The PFSS extrapolation allows an examination of the variation in the large-scale magnetic field of the Sun. The coarse spatial resolution removes the small-scale magnetic field, and thus the complexity of the quiet Sun as well as part of the complexity of active regions. In the following section, the influence of the quiet-Sun magnetic field on the determination of the height of the feature is investigated. 

\subsection{Quiet-Sun model}
\label{SS:qs_model}

\begin{figure*}
\centering
\includegraphics[width=1\linewidth,trim=0 25 0 0, clip=]{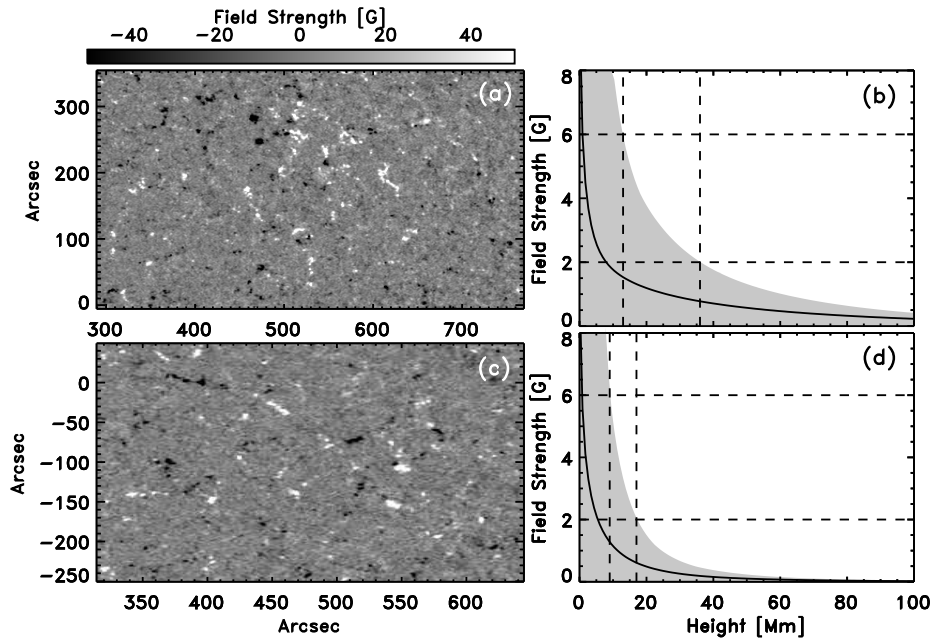}
\caption{\corr{Quiet-Sun magnetograms from 12~June~2010 (panel (a); field-of-view of 480\arcsec$\times$360\arcsec) and 16~February~2011 (panel (c); field-of-view of 330\arcsec$\times$300\arcsec)} used as a boundary condition for the local domain model. \corr{The corresponding average field strength as a function of height in the corona is shown in panel (b) for the event of 12~June~2010 and panel (d) for the event of 16~February~2011}. The grey area shows the spread of the magnetic-field strength. The vertical dashed lines indicate the height at which a maximum field strength of 6~G and 2~G is achieved (horizontal lines).}
\label{fig:b_local}
\end{figure*}

The procedure described by \citet{reg08} is adopted here to compute the potential field above quiet-Sun regions. \corr{SDO}/HMI magnetograms are again employed, although in this case magnetograms closest in time to observations of the ``EIT wave'' are used rather than synoptic maps. The initial pixel size of 0.6\corr{\arcsec} for the \corr{SDO}/HMI magnetograms is increased by a factor of 2 for the extrapolation. A small field-of-view encompassing the \emph{Hinode}/EIS slit was extracted from the full-disk observations, with care taken to ensure that there is no contamination from the stronger magnetic field associated with nearby active regions. The field-of-view used for the potential field extrapolation is shown in Figure~\ref{fig:b_local}(a) for the 12~June~2010 event (480\corr{\arcsec}$\times$360\corr{\arcsec}) and Figure~\ref{fig:b_local}(c) for the 16~February~2011 event (330\corr{\arcsec}$\times$300\corr{\arcsec}). The vertical extension \corr{of the computational domain} is approximately 350\corr{\arcsec} (about 250~Mm) for both computations. 

Similar to the approach followed for the PFSS model, the average magnetic-field strength was estimated at a given height, with the resulting variation with height then plotted (see \corr{Figure~\ref{fig:b_local}(b),(d)}). It is determined that the height from which the magnetic-field strength is above 6~G or 2~G is 13~Mm or 36~Mm for the 12~June~2010 event \corr{(panel (b))} and 9~Mm or 17~Mm for the 16~February~2011 event \corr{(panel (d))}. 

The values obtained for the characteristic height of the ``EIT wave'' pulse using the quiet-Sun field extrapolation are quite low, and not consistent with previous estimates or the heights estimated here using density and temperature scale heights. However, they may be explained by the fact that, even if the model includes the small-scale connectivity and complexity of the magnetic field, the intermediate scale of the active-region magnetic field has been removed.

Given the two different extrapolation techniques outlined above, it may be observed that the PFSS model provides more appropriate results for a large-scale event such as the eruption and propagation of an ``EIT wave''. The active-region scale extrapolation cannot be tackled without removing the small-scale field or the large-scale component with \corr{present} limitations on the field-of-view and pixel size that are reasonably possible to use in extrapolation codes.

\section{Discussion and Conclusions}
\label{S:concs}

The results presented here indicate that the magnetic-field strength of the quiet-solar corona may be determined in-situ using ``EIT waves''. High-cadence images from \corr{SDO}/AIA have been analysed using a semi-automated technique for identifying and tracking ``EIT waves'', thus minimising user bias and allowing a more accurate estimation of the kinematic properties of the disturbance. This approach is complemented by high-resolution spectral observations from \emph{Hinode}/EIS, which allow the \corr{number} density of the corona, through which the pulse was propagating, to be determined. 

Two events from 12~June~2010 and 16~February~2011 are examined here using this rare combination of data: they were both observed at high cadence by \corr{SDO}/AIA and \emph{Hinode}/EIS. However, the magnetic structure of the local corona varies between these cases. The pulse on 12~June~2010 \corr{originated} in a relatively isolated, simple-topology active region. The 16~February~2011 event instead erupted from a very complex active region and was tracked across a region of quiet Sun towards an extended active region of much simpler topology.

The combination of pulse kinematics and density measurements allows the coronal magnetic-field strength to be estimated for the region of the quiet corona through which the pulse propagated. Using this approach, some variability in \corr{number} density and magnetic-field strength is observed for the region of the quiet corona studied, showing a range of  2\,--\,6~G. This variability may play a role in \corr{terms of} the observed properties of the pulse, causing it to decelerate and broaden as it propagates through the randomly structured medium \citep[see, \emph{e.g.},][]{Murawski:2001ab}. These observations \corr{indicate that interpretation of the coronal magnetic-field strength, particularly in quiet coronal regions, is not trivial}.

The estimates derived using this combination of data from \corr{SDO}/AIA and \emph{Hinode}/EIS are then compared to magnetic-field extrapolations of the coronal magnetic field. Data from \corr{SDO}/HMI have been analysed using both a global-scale PFSS extrapolation and a local-scale quiet-Sun magnetic-field extrapolation technique outlined by \citet{reg08}. Using each of these techniques it is possible to estimate the height range corresponding to the magnetic-field strength estimated using the propagating ``EIT wave'' pulse. It is found that the values obtained using the quiet-Sun approach are too low to be considered a realistic estimate of the height range. This indicates that the magnetic-field at the height of the pulse is not dominated by the small-scale magnetic-field anchored in the photosphere below.

The PFSS extrapolation produces height estimates of \corr{$\approx$}~\corr{(}70\,--\,130\corr{)}~Mm, which are most consistent with the height range estimated by \citet{Patsourakos:2009bc} and \citet{Kienreich:2009ab} using quadrature observations of an ``EIT wave'' made by \corr{STEREO}/EUVI. This increasing \corr{number} of observations made using a variety of instruments and techniques suggests that this is the true height range at which ``EIT waves'' propagate. Furthermore, this consistency strongly indicates that ``EIT waves'' are a global phenomenon influenced by global-scale features.

Direct measurements of the coronal magnetic-field are particularly difficult to make, with efforts generally focussed on the variation of the magnetic-field strength in active regions rather than the quiet corona. These results indicate that it is possible to estimate the magnetic-field strength in the solar corona using ``EIT waves'' observed using a combination of broadband and spectroscopic imagers.

%
\begin{acks}
\corr{The authors wish to thank the anonymous referee whose comments improved the article.} \emph{Hinode} is a Japanese mission developed and launched by ISAS/JAXA, collaborating with NAOJ as a domestic partner, NASA and STFC (UK) as international partners. Scientific operation of the \emph{Hinode} mission is conducted by the \emph{Hinode} science team organized at ISAS/JAXA. This team mainly consists of scientists from institutes in the partner countries. Support for the post-launch operation is provided by JAXA and NAOJ (Japan), STFC (UK), NASA, ESA, and NSC (Norway). \corr{SDO}/AIA data are courtesy of NASA/\corr{SDO} and the AIA science team. \corr{The research leading to these results has received funding from the European Commission's Seventh Framework Programme under the grant agreement No. 284461 (eHEROES project).}
\end{acks}

%
 \bibliographystyle{spr-mp-sola}
 \bibliography{sol_phys_bibtex.bib}  

\end{article} 
\end{document}